\documentclass[%
  reprint,
  superscriptaddress,
  amsmath,amssymb,
  pra,
  floatfix,
  longbibliography
]{revtex4-2}

\usepackage[T1]{fontenc}
\usepackage[utf8]{inputenc}
\usepackage{bm}
\usepackage{longtable}
\usepackage[colorlinks=true,allcolors=blue]{hyperref}
\usepackage{orcidlink}

\usepackage{braket}

\begin{document}

%\title{Phase-adjusted realification of a \texorpdfstring{$\mathbb{C}^3$}{C3} Kochen-Specker configuration into \texorpdfstring{$\mathbb{R}^6$}{R6}}
\title{Faithful real embedding of a three-dimensional complex Kochen-Specker configuration}

\author{Andrei Khrennikov\,\orcidlink{0000-0002-9857-0938}}
\email{andrei.khrennikov@lnu.se}
\homepage{https://lnu.se/personal/andrei.khrennikov/}

\affiliation{Institutionen f\"or Matematik, Fakulteten f\"or Teknik,
Linnaeus University, SE-351~95 V\"axj\"o, Sweden}

\author{Karl Svozil\,\orcidlink{0000-0001-6554-2802}}
\email{karl.svozil@tuwien.ac.at}
\homepage{http://tph.tuwien.ac.at/~svozil}

\affiliation{Institute for Theoretical Physics,
TU Wien,
Wiedner Hauptstrasse 8-10/136,
1040 Vienna,  Austria}

\date{\today}

\begin{abstract}
We describe a phase-adjusted realification procedure that embeds any finite set of rays in $\mathbb{C}^3$ into $\mathbb{R}^6$. By assigning an appropriate phase to each ray before applying the standard coordinate-wise map, we can arrange that two rays are orthogonal in $\mathbb{C}^3$ if and only if their images are orthogonal in $\mathbb{R}^6$, so the construction yields a faithful orthogonal representation of the original complex configuration. As a concrete example, we consider the 165 projectively distinct rays used in a $\mathbb{C}^3$ Kochen-Specker configuration obtained from mutually unbiased bases, list these 165 rays explicitly in $\mathbb{C}^3$, and give for each of them its image in $\mathbb{R}^6$ under the canonical realification map. We also note that, because the original 3-element contexts are no longer maximal in $\mathbb{R}^6$, the embedded configuration admits two-valued states even though its realisation with maximal contexts in $\mathbb{C}^3$ is Kochen-Specker uncolourable.
\end{abstract}

\maketitle

\section{A Kochen-Specker Proof for the Nonequivalence of Real and Complex Hilbert Space}

A foundational question in quantum theory is whether the use of complex
numbers is merely a matter of mathematical convenience or a fundamental
feature of the physical formalism~\cite{renou-2021,Weilenmann2025PartialIndependencePRL,Hoffreumon2025NoComplex,Hita2025RealQM,Ying2025FoilComplex,Hardy2012LimitedHolism,Stueckelberg1961RealHilbertII,Bednorz2022OptimalDiscrimination,Aleksandrova2013UniversalQubit,Wootters2016OptimalInformationTransfer,Rudolph2002TwoRebitUniversal,McKague2009SimulatingRealHilbert,Erba2024CompositionRule}.
On the one hand, any finite-dimensional complex Hilbert space $\mathbb{C}^n$ can be
represented as a real Hilbert space $\mathbb{R}^{2n}$ via the usual
``realification'' map that separates real and imaginary parts. This is a
typical instance of the general fact that, under sufficiently permissive
assumptions (allowing a change of dimension and additional structure such as
a complex-structure operator), a real formulation of a complex theory is
always possible. Moreover, for many quantum information processing tasks
such as quantum computation and Bell nonlocality, standard real quantum
theory in this sense is already sufficient to reproduce the operational
predictions of the complex theory
\cite{Aleksandrova2013UniversalQubit,Wootters2016OptimalInformationTransfer,Rudolph2002TwoRebitUniversal,McKague2009SimulatingRealHilbert}.
On the other hand, recent theoretical and experimental work has shown that
if one insists on keeping fixed certain natural structural ingredients
(composition via tensor products, the notion of locality, and the usual
Hilbert-space interpretation of contexts as maximal orthonormal sets), then
there exist correlations and logical structures that are realised in complex
Hilbert spaces but cannot be obtained from any purely real Hilbert-space
model~\cite{renou-2021,Weilenmann2025PartialIndependencePRL,Hoffreumon2025NoComplex,Hita2025RealQM,Ying2025FoilComplex,Bednorz2022OptimalDiscrimination,Erba2024CompositionRule}.

It is therefore useful to distinguish three viewpoints:
(i) Whether quantum theory can be formulated over the reals depends on which
structural assumptions one demands to be preserved.
(ii) For a large class of quantum information tasks, real quantum theory is
operationally sufficient.
(iii) There are also more pragmatic, representation-focused approaches that
encode complex quantum mechanics into a real framework at the price of
introducing a formal structure different from quantum theory~\cite{BarriosHita2025RealQM}.

Against this background, in this paper we present a complementary, Kochen-Specker (KS) type argument on the ``no-go'' side for the logical
inequivalence of three-dimensional real and complex Hilbert-space quantum
theories, within the standard Hilbert-space framework where sharp yes--no
propositions correspond to rays and physical contexts are identified with
maximal sets of mutually orthogonal rays. Rather than working with
inequalities and probabilities, our approach is based on the logical
framework of the Kochen-Specker theorem: we consider a finite set of
rays and their orthogonality relations and derive a direct contradiction
with the existence of classical truth assignments, formalised by
two-valued states. In addition, we exploit a concrete phase-adjusted realification into
$\mathbb{R}^6$, which provides an explicit example of the representation-focused
approach in~(iii).

We focus on a specific KS-type configuration introduced by Cabello~\cite{Cabello2025}
as a threefold extension of the Yu--Oh configuration~\cite{Yu-2012}. As pointed
out by Harding and Salinas Schmeis~\cite{harding2025remarksIJTP} and analysed
in detail by Navara and Svozil~\cite{svozil-2025-MYOCHS}, a key structural
ingredient of this configuration is the existence, in $\mathbb{C}^3$, of a complete
set of mutually unbiased bases (MUBs). Two orthonormal bases are mutually
unbiased if every vector of one basis has equal ``overlap'' with every vector
of the other. The theory of MUBs is a rich subject with deep connections to
quantum information and finite geometry
\cite{Schwinger.60,WooFie,Brierley2009,springerlink:10.1007/978-3-540-24633-6_10,durt,mcnulty2024mutually}.
For the present work, the crucial facts are~\cite{durt,mcnulty2024mutually}:
\begin{enumerate}
  \item In $\mathbb{C}^3$ one can construct a complete set of $D+1=4$ MUBs.
  \item In $\mathbb{R}^3$ it is impossible to find even two mutually unbiased
        orthonormal bases.
\end{enumerate}

The Yu-Oh-Cabello-KS configuration we consider~\cite{Cabello2025} is built
explicitly from the four MUBs in $\mathbb{C}^3$ and consists of 165 rays arranged
into 130 contexts; in Section~\ref{sec:KSconfig} we describe this
configuration in detail.

The present paper has two aims. First, building on the work of Harding and
Salinas Schmeis~\cite{harding2025remarksIJTP} and of Navara and
Svozil~\cite{svozil-2025-MYOCHS}, we recast their observations in a
MUB-based setting by collecting in one place the explicit 165-ray,
130-context KS configuration in $\mathbb{C}^3$ due to Cabello~\cite{Cabello2025}
and explaining, in this concrete example, why its orthogonality relations
cannot be realised by any set of unit vectors in $\mathbb{R}^3$ when contexts are
required to be maximal. Thus, while neither the configuration nor the
impossibility of such a real realisation are new, our treatment provides a
self-contained finite illustration of a logical structure---a set of
propositions and contexts---that exists in three-dimensional complex quantum
theory but not in three-dimensional real quantum theory under the same
interpretation of contexts.

Second, and this is the genuinely new part of the present work, we show that
the same orthogonality hypergraph does admit a faithful orthogonal
representation~\cite{lovasz-79} in $\mathbb{R}^6$: by appropriately choosing the
phase of each ray before applying the standard coordinate-wise separation
into real and imaginary parts, we embed all 165 rays as vectors in $\mathbb{R}^6$
so that two rays are orthogonal in $\mathbb{C}^3$ if and only if their images are
orthogonal in $\mathbb{R}^6$. However, in this real embedding each 3-element
context now spans only a 3-dimensional subspace of $\mathbb{R}^6$ and is no longer
maximal. It can always be extended to a full orthonormal basis of $\mathbb{R}^6$
by adding three extra vectors outside the configuration. This allows one to
construct two-valued states by assigning value $0$ to all 165 embedded rays
and value $1$ to a suitable extra basis vector in each completed context.

In this way, the same finite orthogonality hypergraph, realised as maximal
contexts in $\mathbb{C}^3$, is KS-uncolourable and admits no classical two-valued
states, whereas its faithful realisation in $\mathbb{R}^6$ does admit classical
two-valued states once contexts are interpreted as maximal in the larger
space. Our analysis therefore highlights a nuanced picture: at fixed
dimension $3$, and with the usual identification of contexts with maximal
orthonormal sets, there exist logical structures realisable in complex but
not in real quantum theory, in line with earlier work
\cite{harding2025remarksIJTP,svozil-2025-MYOCHS}. At the same time, our
phase-adjusted realification shows that \emph{orthogonality relations
alone} do not separate real from complex quantum theory, since any finite
orthogonality hypergraph arising from rays in $\mathbb{C}^3$ admits a faithful
orthogonal representation in a suitable real Hilbert space. This clarifies
the scope of KS-style arguments in the real--complex debate and complements
earlier correlation-based separations between real and complex
Hilbert-space models.

Our argument is, in this sense, system-dependent but basis-independent: it
singles out a specific three-dimensional configuration and shows that it can
be realised in complex but not in real Hilbert space of the same dimension
when contexts are required to be maximal, yet the statement does not depend
on any preferred basis. This positions our work between basis- and
system-dependent, representation-focused approaches (such as resource
theories of coherence or imaginarity) and more global, semi-device-independent
separations between real and complex quantum theories in the style of
Renou \emph{et al.}~\cite{renou-2021}.

\section{Phase-adjusted realification in \texorpdfstring{$\mathbb{C}^3$}{C3}}

\subsection{Set-up}

Let $\{\,[\psi_k]\,\}_{k=1}^N$ be a finite set of rays in $\mathbb{C}^3$, with normalized representatives $\psi_k\in\mathbb{C}^3$, and let
\begin{equation}
  c_{k\ell} := \langle \psi_k,\psi_\ell\rangle_{\mathbb{C}}
\end{equation}
denote their complex inner products.

For each ray we may multiply $\psi_k$ by an arbitrary unit complex phase
$e^{\mathrm{i}\theta_k}$ without changing the ray. We first define the canonical realification map $\Phi_0: \mathbb{C}^3 \to \mathbb{R}^6$ as
\begin{equation}
  \Phi_0(z_1, z_2, z_3) := (\Re z_1, \Re z_2, \Re z_3, \Im z_1, \Im z_2, \Im z_3).
  \label{eq:phi0}
\end{equation}
In the physics literature this
invariance under multiplication by a global phase $e^{\mathrm{i}\theta_k}\in U(1)$
is often referred to as an (abelian) gauge freedom. We exploit this freedom
and define a \emph{phase-adjusted realification}
\begin{equation}
  R_k := \Phi_0\bigl( e^{\mathrm{i}\theta_k}\,\psi_k\bigr)\in\mathbb{R}^6,
\label{2025-rvc-padef}
\end{equation}

Note that the real part of the complex inner product is exactly the real dot
product:
\begin{align}
  R_k\cdot R_\ell
  &= \Phi_0(e^{\mathrm{i}\theta_k}\psi_k)\cdot \Phi_0(e^{\mathrm{i}\theta_\ell}\psi_\ell) \nonumber\\
  &= \Re\Bigl\langle e^{\mathrm{i}\theta_k}\psi_k, e^{\mathrm{i}\theta_\ell}\psi_\ell\Bigr\rangle_{\mathbb{C}} \nonumber\\
  &= \Re\bigl( e^{\mathrm{i}(\theta_\ell-\theta_k)} c_{k\ell}\bigr).
  \label{eq:real-dot}
\end{align}
If $c_{k\ell}=0$, then $R_k\cdot R_\ell=0$ for all phases, so complex orthogonality is always preserved. The only potential issue is that for $c_{k\ell}\neq 0$ we might accidentally choose phases such that $R_k\cdot R_\ell=0$, thereby introducing a spurious orthogonality.

\subsection{Forbidden phase differences for non-orthogonal pairs}

Assume $c_{k\ell}\neq 0$. Write $c_{k\ell}=|c_{k\ell}|e^{\mathrm{i}\varphi_{k\ell}}$. Then Eq.~\eqref{eq:real-dot} becomes
\begin{equation}
  R_k\cdot R_\ell
  = |c_{k\ell}|\,\cos\bigl( (\theta_\ell-\theta_k)+\varphi_{k\ell}\bigr).
\end{equation}
Thus $R_k\cdot R_\ell=0$ if and only if
\begin{equation}
  (\theta_\ell-\theta_k)+\varphi_{k\ell} \equiv \frac{\pi}{2}\pmod{\pi}.
\end{equation}
For a fixed $\theta_k$ and fixed nonzero $c_{k\ell}$, this excludes at most two values of $\theta_\ell$ modulo $2\pi$:
\begin{equation}
  \theta_\ell \equiv \theta_k -\varphi_{k\ell} + \frac{\pi}{2}
  \quad\text{or}\quad
  \theta_\ell \equiv \theta_k -\varphi_{k\ell} + \frac{3\pi}{2}
  \pmod{2\pi}.
\label{2025-rvc-pa}
\end{equation}

Changing $\theta_k$ shifts both forbidden values of $\theta_\ell$ by the same amount, reflecting the overall phase (gauge) freedom in the choice of ray representatives.

\subsection{Existence of a faithful embedding}

We construct the phases $\theta_1,\dots,\theta_N$ inductively.

\begin{itemize}
  \item Set $\theta_1:=0$ arbitrarily.
  \item Suppose $\theta_1,\dots,\theta_{m-1}$ have been chosen such that for all non-orthogonal pairs $(k,\ell)$ with $1\le k<\ell<m$ we have $R_k\cdot R_\ell\neq 0$.
  \item For the new ray $m$, consider all pairs $(k,m)$ with $1\le k<m$ and $c_{km}\neq 0$. For each such pair, the two forbidden values of $\theta_m$ described above define a finite set $F_{k,m}\subset S^1$, where $S^1$ denotes the unit circle.
\end{itemize}

Let
\begin{equation}
  F_m := \bigcup_{\substack{1\le k<m \\ c_{km}\neq 0}} F_{k,m},
\end{equation}
which is a finite subset of the unit circle. Choose $\theta_m$ anywhere on $S^1\setminus F_m$---that is, from the set of admissible phase values excluding the finite forbidden set $F_m$.
By construction, for every non-orthogonal pair $(k,m)$ we have $R_k\cdot R_m\neq 0$, while pairs with $c_{km}=0$ remain orthogonal for all phases. Since the forbidden set $F_m$ is finite and the circle is connected, such a $\theta_m$ always exists; in fact, all but finitely many phases work. Proceeding by induction on $m$, we can successively choose the phases $\theta_m$ so that no unwanted orthogonality is introduced, obtaining a full set of phases $\{\theta_k\}_{k=1}^N$ such that
\begin{equation}
  c_{k\ell}=0 \iff R_k\cdot R_\ell=0.
\end{equation}
Thus $\{R_k\}_{k=1}^N\subset\mathbb{R}^6$ is a faithful orthogonal representation of the original finite set of rays in $\mathbb{C}^3$.

In particular, if $\{\psi_{i},\psi_{j},\psi_{k}\}$ form an orthonormal basis of $\mathbb{C}^3$, then each pair is orthogonal in $\mathbb{C}^3$, hence each pair $R_i,R_j,R_k$ is orthogonal in $\mathbb{R}^6$, so orthogonal triples (contexts) are preserved.

\section{Explicit list of the 165 rays and their images in \texorpdfstring{$\mathbb{R}^6$}{R6}}
\label{sec:KSconfig}

\subsection{Complex notation}

In the previous section we showed in general that any finite family of rays in $\mathbb{C}^3$
admits a faithful orthogonal representation in $\mathbb{R}^6$ by an appropriate choice of
phases and the realification map $\Phi_0$. We now specialise this construction to the
concrete KS-type configuration introduced by Cabello and described in
Sect.~\ref{sec:KSconfig}: in this section we fix explicit representatives for the 165
rays and write down their images in $\mathbb{R}^6$.

To keep the expressions compact, it is convenient to exploit the algebraic structure of
this configuration. Since it is built from four mutually unbiased bases in dimension
three, all vector coordinates lie in the cyclotomic field generated by a primitive
third root of unity. We therefore introduce the shorthand
\begin{equation}
  \omega := e^{2\pi\mathrm{i}/3} = -\frac12 + \mathrm{i}\,\frac{\sqrt{3}}{2}, \qquad
  \omega^2 = \overline{\omega} = -\frac12 - \mathrm{i}\,\frac{\sqrt{3}}{2},
\end{equation}
with $1+\omega+\omega^2=0$.

Every coordinate in the 165 rays below is a real multiple of $1$, $\omega$, or $\omega^2$ with integer coefficients in $\{-2,-1,0,1,2\}$. For any $z=a+b\,\omega+c\,\omega^2$ ($a,b,c\in\mathbb{R}$) we have
\begin{align}
  \Re z &= a - \frac{b+c}{2},\\
  \Im z &= \frac{\sqrt{3}}{2}\,(b-c).
\end{align}
In particular, $0 \mapsto (0,0)$, and
\begin{align}
  1 &\mapsto (1,0), &
  -1 &\mapsto (-1,0), \nonumber\\
  \omega &\mapsto \Bigl(-\tfrac12,\ \tfrac{\sqrt{3}}{2}\Bigr), &
  \omega^2 &\mapsto \Bigl(-\tfrac12,\ -\tfrac{\sqrt{3}}{2}\Bigr), \nonumber\\
  2\omega &\mapsto \bigl(-1,\ \sqrt{3}\bigr), &
  2\omega^2 &\mapsto \bigl(-1,\ -\sqrt{3}\bigr), \nonumber\\
  -\omega &\mapsto \Bigl(\tfrac12,\ -\tfrac{\sqrt{3}}{2}\Bigr), &
  -\omega^2 &\mapsto \Bigl(\tfrac12,\ \tfrac{\sqrt{3}}{2}\Bigr), \\
  -2\omega &\mapsto \bigl(1,\ -\sqrt{3}\bigr), &
  -2\omega^2 &\mapsto \bigl(1,\ \sqrt{3}\bigr). \nonumber
\end{align}

Since three complex components are encoded as six real ones, the resulting
$\mathbb{R}^6$ representation is not unique: any permutation of the six real
coordinates gives an orthogonally equivalent configuration. In particular,
there are $3!\cdot 2^3 = 48$ permutations that merely reshuffle the three
complex entries and, for each entry, possibly swap its real and imaginary
part. For example, the vector $(1,1,1,\,0,0,0)$ is permutationally
equivalent to $(1,0,1,0,1,0)$ via the permutation $(2\,4\,5\,3)$ on
$\{1,\dots,6\}$; that is,
\[
(x_1,x_2,x_3,x_4,x_5,x_6)\mapsto(x_1,x_4,x_2,x_5,x_3,x_6).
\]
More generally, composing our phase-adjusted realification with any real
orthogonal transformation of $\mathbb{R}^6$ yields another faithful orthogonal
representation of the same 165-ray configuration: all such choices are
isomorphic from the point of view of the orthogonality hypergraph and of
KS-colourability. In the table below we fix a particularly simple
``canonical gauge'', namely the ordering
$(\Re z_1,\Re z_2,\Re z_3,\Im z_1,\Im z_2,\Im z_3)$, but none of our
conclusions depend on this specific coordinate choice.

\subsection{Explicit realification}
\label{sec:explicit_realification}

For concreteness, in Table~\ref{tab:all-vectors} we provide the explicit coordinates for all 165 rays in the configuration. The orthogonality graph of these rays---which visualises the logical structure of the proof---is depicted in Fig.~2 of Ref.~\cite{harding2025remarksIJTP}, as well as in Fig.~1 of Ref.~\cite{Cabello2025} and Fig.~1 of Ref.~\cite{svozil-2025-MYOCHS}. In our tabulation, each row contains a label, the corresponding ray in $\mathbb{C}^3$, and its image in $\mathbb{R}^6$ under the canonical realification map~\eqref{eq:phi0} (i.e., with phases $\theta_k=0$). For a vector $(z_1,z_2,z_3)\in\mathbb{C}^3$ we write its real image as
\begin{equation}
  (\Re z_1,\Re z_2,\Re z_3,\Im z_1,\Im z_2,\Im z_3)\in\mathbb{R}^6.
\end{equation}

\begingroup
\renewcommand{\arraystretch}{1.5} % Increases vertical spacing for fractions
\begin{longtable}{lll}
\caption{All 165 rays in $\mathbb{C}^3$ and their images in $\mathbb{R}^6$
under the canonical realification~\eqref{eq:phi0} (i.e. with all phase adjustments $\theta_k=0 $ as defined in~\eqref{2025-rvc-padef}).
This brings about spurious orthogonalities in $\mathbb{R}^6$, such as between $u_{1}$ and $u_{4}$ that are not present in $\mathbb{C}^3$. A strictly faithful embedding with no spurious orthogonalities requires the additional phase adjustments described in~\eqref{2025-rvc-pa}.}
\label{tab:all-vectors}\\
\hline\hline
Label & $\mathbb{C}^3$ vector & $\mathbb{R}^6$ vector \\
\hline
\endfirsthead
\hline\hline
Label & $\mathbb{C}^3$ vector & $\mathbb{R}^6$ vector \\
\hline
\endhead
\hline\hline
\endfoot
$a_{11}$ & $(1,0,0)$ &
$\bigl(1,0,0,\,0,0,0\bigr)$ \\
$a_{21}$ & $(0,1,0)$ &
$\bigl(0,1,0,\,0,0,0\bigr)$ \\
$a_{31}$ & $(0,0,1)$ &
$\bigl(0,0,1,\,0,0,0\bigr)$ \\
$u_{1}$ & $(1,1,1)$ &
$\bigl(1,1,1,\,0,0,0\bigr)$ \\
$u_{2}$ & $(1,\omega,\omega^2)$ &
$\Bigl(1,-\tfrac12,-\tfrac12,\,0,\tfrac{\sqrt{3}}{2},-\tfrac{\sqrt{3}}{2}\Bigr)$ \\
$u_{3}$ & $(1,\omega^2,\omega)$ &
$\Bigl(1,-\tfrac12,-\tfrac12,\,0,-\tfrac{\sqrt{3}}{2},\tfrac{\sqrt{3}}{2}\Bigr)$ \\
$u_{4}$ & $(1,\omega,\omega)$ &
$\Bigl(1,-\tfrac12,-\tfrac12,\,0,\tfrac{\sqrt{3}}{2},\tfrac{\sqrt{3}}{2}\Bigr)$ \\
$u_{5}$ & $(1,\omega^2,1)$ &
$\Bigl(1,-\tfrac12,1,\,0,-\tfrac{\sqrt{3}}{2},0\Bigr)$ \\
$u_{6}$ & $(1,1,\omega^2)$ &
$\Bigl(1,1,-\tfrac12,\,0,0,-\tfrac{\sqrt{3}}{2}\Bigr)$ \\
$u_{7}$ & $(1,\omega^2,\omega^2)$ &
$\Bigl(1,-\tfrac12,-\tfrac12,\,0,-\tfrac{\sqrt{3}}{2},-\tfrac{\sqrt{3}}{2}\Bigr)$ \\
$u_{8}$ & $(1,\omega,1)$ &
$\Bigl(1,-\tfrac12,1,\,0,\tfrac{\sqrt{3}}{2},0\Bigr)$ \\
$u_{9}$ & $(1,1,\omega)$ &
$\Bigl(1,1,-\tfrac12,\,0,0,\tfrac{\sqrt{3}}{2}\Bigr)$ \\
$b_{11}$ & $(0,1,1)$ &
$\bigl(0,1,1,\,0,0,0\bigr)$ \\
$b_{12}$ & $(0,\omega,\omega^2)$ &
$\Bigl(0,-\tfrac12,-\tfrac12,\,0,\tfrac{\sqrt{3}}{2},-\tfrac{\sqrt{3}}{2}\Bigr)$ \\
$b_{13}$ & $(0,\omega^2,\omega)$ &
$\Bigl(0,-\tfrac12,-\tfrac12,\,0,-\tfrac{\sqrt{3}}{2},\tfrac{\sqrt{3}}{2}\Bigr)$ \\
$b_{21}$ & $(1,0,1)$ &
$\bigl(1,0,1,\,0,0,0\bigr)$ \\
$b_{22}$ & $(1,0,\omega^2)$ &
$\Bigl(1,0,-\tfrac12,\,0,0,-\tfrac{\sqrt{3}}{2}\Bigr)$ \\
$b_{23}$ & $(1,0,\omega)$ &
$\Bigl(1,0,-\tfrac12,\,0,0,\tfrac{\sqrt{3}}{2}\Bigr)$ \\
$b_{31}$ & $(1,1,0)$ &
$\bigl(1,1,0,\,0,0,0\bigr)$ \\
$b_{32}$ & $(1,\omega,0)$ &
$\Bigl(1,-\tfrac12,0,\,0,\tfrac{\sqrt{3}}{2},0\Bigr)$ \\
$b_{33}$ & $(1,\omega^2,0)$ &
$\Bigl(1,-\tfrac12,0,\,0,-\tfrac{\sqrt{3}}{2},0\Bigr)$ \\
$c_{11}$ & $(0,1,-1)$ &
$\bigl(0,1,-1,\,0,0,0\bigr)$ \\
$c_{12}$ & $(0,\omega,-\omega^2)$ &
$\Bigl(0,-\tfrac12,\tfrac12,\,0,\tfrac{\sqrt{3}}{2},\tfrac{\sqrt{3}}{2}\Bigr)$ \\
$c_{13}$ & $(0,\omega^2,-\omega)$ &
$\Bigl(0,-\tfrac12,\tfrac12,\,0,-\tfrac{\sqrt{3}}{2},-\tfrac{\sqrt{3}}{2}\Bigr)$ \\
$c_{21}$ & $(-1,0,1)$ &
$\bigl(-1,0,1,\,0,0,0\bigr)$ \\
$c_{22}$ & $(-\omega,0,1)$ &
$\Bigl(\tfrac12,0,1,\,-\tfrac{\sqrt{3}}{2},0,0\Bigr)$ \\
$c_{23}$ & $(-\omega^2,0,1)$ &
$\Bigl(\tfrac12,0,1,\,\tfrac{\sqrt{3}}{2},0,0\Bigr)$ \\
$c_{31}$ & $(1,-1,0)$ &
$\bigl(1,-1,0,\,0,0,0\bigr)$ \\
$c_{32}$ & $(\omega^2,-1,0)$ &
$\Bigl(-\tfrac12,-1,0,\,-\tfrac{\sqrt{3}}{2},0,0\Bigr)$ \\
$c_{33}$ & $(\omega,-1,0)$ &
$\Bigl(-\tfrac12,-1,0,\,\tfrac{\sqrt{3}}{2},0,0\Bigr)$ \\
$d_{11}$ & $(-2,1,1)$ &
$\bigl(-2,1,1,\,0,0,0\bigr)$ \\
$d_{12}$ & $(-2,\omega,\omega^2)$ &
$\Bigl(-2,-\tfrac12,-\tfrac12,\,0,\tfrac{\sqrt{3}}{2},-\tfrac{\sqrt{3}}{2}\Bigr)$ \\
$d_{13}$ & $(-2,\omega^2,\omega)$ &
$\Bigl(-2,-\tfrac12,-\tfrac12,\,0,-\tfrac{\sqrt{3}}{2},\tfrac{\sqrt{3}}{2}\Bigr)$ \\
$d_{14}$ & $(-2,\omega,\omega)$ &
$\Bigl(-2,-\tfrac12,-\tfrac12,\,0,\tfrac{\sqrt{3}}{2},\tfrac{\sqrt{3}}{2}\Bigr)$ \\
$d_{15}$ & $(-2,\omega^2,1)$ &
$\Bigl(-2,-\tfrac12,1,\,0,-\tfrac{\sqrt{3}}{2},0\Bigr)$ \\
$d_{16}$ & $(-2,1,\omega^2)$ &
$\Bigl(-2,1,-\tfrac12,\,0,0,-\tfrac{\sqrt{3}}{2}\Bigr)$ \\
$d_{17}$ & $(-2,\omega^2,\omega^2)$ &
$\Bigl(-2,-\tfrac12,-\tfrac12,\,0,-\tfrac{\sqrt{3}}{2},-\tfrac{\sqrt{3}}{2}\Bigr)$ \\
$d_{18}$ & $(-2,\omega,1)$ &
$\Bigl(-2,-\tfrac12,1,\,0,\tfrac{\sqrt{3}}{2},0\Bigr)$ \\
$d_{19}$ & $(-2,1,\omega)$ &
$\Bigl(-2,1,-\tfrac12,\,0,0,\tfrac{\sqrt{3}}{2}\Bigr)$ \\
$d_{21}$ & $(1,-2,1)$ &
$\bigl(1,-2,1,\,0,0,0\bigr)$ \\
$d_{22}$ & $(\omega^2,-2,\omega)$ &
$\Bigl(-\tfrac12,-2,-\tfrac12,\,-\tfrac{\sqrt{3}}{2},0,\tfrac{\sqrt{3}}{2}\Bigr)$ \\
$d_{23}$ & $(\omega,-2,\omega^2)$ &
$\Bigl(-\tfrac12,-2,-\tfrac12,\,\tfrac{\sqrt{3}}{2},0,-\tfrac{\sqrt{3}}{2}\Bigr)$ \\
$d_{24}$ & $(\omega^2,-2,1)$ &
$\Bigl(-\tfrac12,-2,1,\,-\tfrac{\sqrt{3}}{2},0,0\Bigr)$ \\
$d_{25}$ & $(\omega,-2,\omega)$ &
$\Bigl(-\tfrac12,-2,-\tfrac12,\,\tfrac{\sqrt{3}}{2},0,\tfrac{\sqrt{3}}{2}\Bigr)$ \\
$d_{26}$ & $(1,-2,\omega^2)$ &
$\Bigl(1,-2,-\tfrac12,\,0,0,-\tfrac{\sqrt{3}}{2}\Bigr)$ \\
$d_{27}$ & $(\omega,-2,1)$ &
$\Bigl(-\tfrac12,-2,1,\,\tfrac{\sqrt{3}}{2},0,0\Bigr)$ \\
$d_{28}$ & $(\omega^2,-2,\omega^2)$ &
$\Bigl(-\tfrac12,-2,-\tfrac12,\,-\tfrac{\sqrt{3}}{2},0,-\tfrac{\sqrt{3}}{2}\Bigr)$ \\
$d_{29}$ & $(1,-2,\omega)$ &
$\Bigl(1,-2,-\tfrac12,\,0,0,\tfrac{\sqrt{3}}{2}\Bigr)$ \\
$d_{31}$ & $(1,1,-2)$ &
$\bigl(1,1,-2,\,0,0,0\bigr)$ \\
$d_{32}$ & $(\omega,\omega^2,-2)$ &
$\Bigl(-\tfrac12,-\tfrac12,-2,\,\tfrac{\sqrt{3}}{2},-\tfrac{\sqrt{3}}{2},0\Bigr)$ \\
$d_{33}$ & $(\omega^2,\omega,-2)$ &
$\Bigl(-\tfrac12,-\tfrac12,-2,\,-\tfrac{\sqrt{3}}{2},\tfrac{\sqrt{3}}{2},0\Bigr)$ \\
$d_{34}$ & $(\omega^2,1,-2)$ &
$\Bigl(-\tfrac12,1,-2,\,-\tfrac{\sqrt{3}}{2},0,0\Bigr)$ \\
$d_{35}$ & $(1,\omega^2,-2)$ &
$\Bigl(1,-\tfrac12,-2,\,0,-\tfrac{\sqrt{3}}{2},0\Bigr)$ \\
$d_{36}$ & $(\omega,\omega,-2)$ &
$\Bigl(-\tfrac12,-\tfrac12,-2,\,\tfrac{\sqrt{3}}{2},\tfrac{\sqrt{3}}{2},0\Bigr)$ \\
$d_{37}$ & $(\omega,1,-2)$ &
$\Bigl(-\tfrac12,1,-2,\,\tfrac{\sqrt{3}}{2},0,0\Bigr)$ \\
$d_{38}$ & $(1,\omega,-2)$ &
$\Bigl(1,-\tfrac12,-2,\,0,\tfrac{\sqrt{3}}{2},0\Bigr)$ \\
$d_{39}$ & $(\omega^2,\omega^2,-2)$ &
$\Bigl(-\tfrac12,-\tfrac12,-2,\,-\tfrac{\sqrt{3}}{2},-\tfrac{\sqrt{3}}{2},0\Bigr)$ \\
$b_{121}$ & $(1,1,-1)$ &
$\bigl(1,1,-1,\,0,0,0\bigr)$ \\
$b_{122}$ & $(1,\omega,-\omega^2)$ &
$\Bigl(1,-\tfrac12,\tfrac12,\,0,\tfrac{\sqrt{3}}{2},\tfrac{\sqrt{3}}{2}\Bigr)$ \\
$b_{123}$ & $(1,\omega^2,-\omega)$ &
$\Bigl(1,-\tfrac12,\tfrac12,\,0,-\tfrac{\sqrt{3}}{2},-\tfrac{\sqrt{3}}{2}\Bigr)$ \\
$b_{124}$ & $(\omega,\omega^2,-\omega^2)$ &
$\Bigl(-\tfrac12,-\tfrac12,\tfrac12,\,\tfrac{\sqrt{3}}{2},-\tfrac{\sqrt{3}}{2},\tfrac{\sqrt{3}}{2}\Bigr)$ \\
$b_{125}$ & $(\omega,1,-\omega)$ &
$\Bigl(-\tfrac12,1,\tfrac12,\,\tfrac{\sqrt{3}}{2},0,-\tfrac{\sqrt{3}}{2}\Bigr)$ \\
$b_{126}$ & $(\omega,\omega,-1)$ &
$\Bigl(-\tfrac12,-\tfrac12,-1,\,\tfrac{\sqrt{3}}{2},\tfrac{\sqrt{3}}{2},0\Bigr)$ \\
$b_{127}$ & $(\omega^2,\omega,-\omega)$ &
$\Bigl(-\tfrac12,-\tfrac12,\tfrac12,\,-\tfrac{\sqrt{3}}{2},\tfrac{\sqrt{3}}{2},-\tfrac{\sqrt{3}}{2}\Bigr)$ \\
$b_{128}$ & $(\omega^2,1,-\omega^2)$ &
$\Bigl(-\tfrac12,1,\tfrac12,\,-\tfrac{\sqrt{3}}{2},0,\tfrac{\sqrt{3}}{2}\Bigr)$ \\
$b_{129}$ & $(\omega^2,\omega^2,-1)$ &
$\Bigl(-\tfrac12,-\tfrac12,-1,\,-\tfrac{\sqrt{3}}{2},-\tfrac{\sqrt{3}}{2},0\Bigr)$ \\
$b_{131}$ & $(1,-1,1)$ &
$\bigl(1,-1,1,\,0,0,0\bigr)$ \\
$b_{132}$ & $(1,-\omega,\omega^2)$ &
$\Bigl(1,\tfrac12,-\tfrac12,\,0,-\tfrac{\sqrt{3}}{2},-\tfrac{\sqrt{3}}{2}\Bigr)$ \\
$b_{133}$ & $(1,-\omega^2,\omega)$ &
$\Bigl(1,\tfrac12,-\tfrac12,\,0,\tfrac{\sqrt{3}}{2},\tfrac{\sqrt{3}}{2}\Bigr)$ \\
$b_{134}$ & $(\omega,-\omega^2,\omega^2)$ &
$\Bigl(-\tfrac12,\tfrac12,-\tfrac12,\,\tfrac{\sqrt{3}}{2},\tfrac{\sqrt{3}}{2},-\tfrac{\sqrt{3}}{2}\Bigr)$ \\
$b_{135}$ & $(\omega,-1,\omega)$ &
$\Bigl(-\tfrac12,-1,-\tfrac12,\,\tfrac{\sqrt{3}}{2},0,\tfrac{\sqrt{3}}{2}\Bigr)$ \\
$b_{136}$ & $(\omega,-\omega,1)$ &
$\Bigl(-\tfrac12,\tfrac12,1,\,\tfrac{\sqrt{3}}{2},-\tfrac{\sqrt{3}}{2},0\Bigr)$ \\
$b_{137}$ & $(\omega^2,-\omega,\omega)$ &
$\Bigl(-\tfrac12,\tfrac12,-\tfrac12,\,-\tfrac{\sqrt{3}}{2},-\tfrac{\sqrt{3}}{2},\tfrac{\sqrt{3}}{2}\Bigr)$ \\
$b_{138}$ & $(\omega^2,-1,\omega^2)$ &
$\Bigl(-\tfrac12,-1,-\tfrac12,\,-\tfrac{\sqrt{3}}{2},0,-\tfrac{\sqrt{3}}{2}\Bigr)$ \\
$b_{139}$ & $(\omega^2,-\omega^2,1)$ &
$\Bigl(-\tfrac12,\tfrac12,1,\,-\tfrac{\sqrt{3}}{2},\tfrac{\sqrt{3}}{2},0\Bigr)$ \\
$b_{231}$ & $(-1,1,1)$ &
$\bigl(-1,1,1,\,0,0,0\bigr)$ \\
$b_{232}$ & $(-1,\omega,\omega^2)$ &
$\Bigl(-1,-\tfrac12,-\tfrac12,\,0,\tfrac{\sqrt{3}}{2},-\tfrac{\sqrt{3}}{2}\Bigr)$ \\
$b_{233}$ & $(-1,\omega^2,\omega)$ &
$\Bigl(-1,-\tfrac12,-\tfrac12,\,0,-\tfrac{\sqrt{3}}{2},\tfrac{\sqrt{3}}{2}\Bigr)$ \\
$b_{234}$ & $(-\omega,\omega^2,\omega^2)$ &
$\Bigl(\tfrac12,-\tfrac12,-\tfrac12,\,-\tfrac{\sqrt{3}}{2},-\tfrac{\sqrt{3}}{2},-\tfrac{\sqrt{3}}{2}\Bigr)$ \\
$b_{235}$ & $(-\omega,1,\omega)$ &
$\Bigl(\tfrac12,1,-\tfrac12,\,-\tfrac{\sqrt{3}}{2},0,\tfrac{\sqrt{3}}{2}\Bigr)$ \\
$b_{236}$ & $(-\omega,\omega,1)$ &
$\Bigl(\tfrac12,-\tfrac12,1,\,-\tfrac{\sqrt{3}}{2},\tfrac{\sqrt{3}}{2},0\Bigr)$ \\
$b_{237}$ & $(-\omega^2,\omega,\omega)$ &
$\Bigl(\tfrac12,-\tfrac12,-\tfrac12,\,\tfrac{\sqrt{3}}{2},\tfrac{\sqrt{3}}{2},\tfrac{\sqrt{3}}{2}\Bigr)$ \\
$b_{238}$ & $(-\omega^2,1,\omega^2)$ &
$\Bigl(\tfrac12,1,-\tfrac12,\,\tfrac{\sqrt{3}}{2},0,-\tfrac{\sqrt{3}}{2}\Bigr)$ \\
$b_{239}$ & $(-\omega^2,\omega^2,1)$ &
$\Bigl(\tfrac12,-\tfrac12,1,\,\tfrac{\sqrt{3}}{2},-\tfrac{\sqrt{3}}{2},0\Bigr)$ \\
$e_{11}$ & $(1,2,1)$ &
$\bigl(1,2,1,\,0,0,0\bigr)$ \\
$e_{12}$ & $(\omega^2,2,\omega)$ &
$\Bigl(-\tfrac12,2,-\tfrac12,\,-\tfrac{\sqrt{3}}{2},0,\tfrac{\sqrt{3}}{2}\Bigr)$ \\
$e_{13}$ & $(\omega,2,\omega^2)$ &
$\Bigl(-\tfrac12,2,-\tfrac12,\,\tfrac{\sqrt{3}}{2},0,-\tfrac{\sqrt{3}}{2}\Bigr)$ \\
$e_{14}$ & $(\omega,2\omega^2,\omega^2)$ &
$\Bigl(-\tfrac12,-1,-\tfrac12,\,\tfrac{\sqrt{3}}{2},-\sqrt{3},-\tfrac{\sqrt{3}}{2}\Bigr)$ \\
$e_{15}$ & $(1,2\omega^2,1)$ &
$\bigl(1,-1,1,\,0,-\sqrt{3},0\bigr)$ \\
$e_{16}$ & $(\omega^2,2\omega^2,\omega)$ &
$\Bigl(-\tfrac12,-1,-\tfrac12,\,-\tfrac{\sqrt{3}}{2},-\sqrt{3},\tfrac{\sqrt{3}}{2}\Bigr)$ \\
$e_{17}$ & $(\omega^2,2\omega,\omega)$ &
$\Bigl(-\tfrac12,-1,-\tfrac12,\,-\tfrac{\sqrt{3}}{2},\sqrt{3},\tfrac{\sqrt{3}}{2}\Bigr)$ \\
$e_{18}$ & $(1,2\omega,1)$ &
$\bigl(1,-1,1,\,0,\sqrt{3},0\bigr)$ \\
$e_{19}$ & $(\omega,2\omega,\omega^2)$ &
$\Bigl(-\tfrac12,-1,-\tfrac12,\,\tfrac{\sqrt{3}}{2},\sqrt{3},-\tfrac{\sqrt{3}}{2}\Bigr)$ \\
$e_{21}$ & $(1,1,2)$ &
$\bigl(1,1,2,\,0,0,0\bigr)$ \\
$e_{22}$ & $(\omega,\omega^2,2)$ &
$\Bigl(-\tfrac12,-\tfrac12,2,\,\tfrac{\sqrt{3}}{2},-\tfrac{\sqrt{3}}{2},0\Bigr)$ \\
$e_{23}$ & $(\omega^2,\omega,2)$ &
$\Bigl(-\tfrac12,-\tfrac12,2,\,-\tfrac{\sqrt{3}}{2},\tfrac{\sqrt{3}}{2},0\Bigr)$ \\
$e_{24}$ & $(\omega,\omega^2,2\omega^2)$ &
$\Bigl(-\tfrac12,-\tfrac12,-1,\,\tfrac{\sqrt{3}}{2},-\tfrac{\sqrt{3}}{2},-\sqrt{3}\Bigr)$ \\
$e_{25}$ & $(\omega^2,\omega,2\omega^2)$ &
$\Bigl(-\tfrac12,-\tfrac12,-1,\,-\tfrac{\sqrt{3}}{2},\tfrac{\sqrt{3}}{2},-\sqrt{3}\Bigr)$ \\
$e_{26}$ & $(1,1,2\omega^2)$ &
$\bigl(1,1,-1,\,0,0,-\sqrt{3}\bigr)$ \\
$e_{27}$ & $(\omega^2,\omega,2\omega)$ &
$\Bigl(-\tfrac12,-\tfrac12,-1,\,-\tfrac{\sqrt{3}}{2},\tfrac{\sqrt{3}}{2},\sqrt{3}\Bigr)$ \\
$e_{28}$ & $(\omega,\omega^2,2\omega)$ &
$\Bigl(-\tfrac12,-\tfrac12,-1,\,\tfrac{\sqrt{3}}{2},-\tfrac{\sqrt{3}}{2},\sqrt{3}\Bigr)$ \\
$e_{29}$ & $(1,1,2\omega)$ &
$\bigl(1,1,-1,\,0,0,\sqrt{3}\bigr)$ \\
$e_{31}$ & $(2,1,1)$ &
$\bigl(2,1,1,\,0,0,0\bigr)$ \\
$e_{32}$ & $(2,\omega,\omega^2)$ &
$\Bigl(2,-\tfrac12,-\tfrac12,\,0,\tfrac{\sqrt{3}}{2},-\tfrac{\sqrt{3}}{2}\Bigr)$ \\
$e_{33}$ & $(2,\omega^2,\omega)$ &
$\Bigl(2,-\tfrac12,-\tfrac12,\,0,-\tfrac{\sqrt{3}}{2},\tfrac{\sqrt{3}}{2}\Bigr)$ \\
$e_{34}$ & $(2\omega^2,1,1)$ &
$\bigl(-1,1,1,\,-\sqrt{3},0,0\bigr)$ \\
$e_{35}$ & $(2\omega^2,\omega,\omega^2)$ &
$\Bigl(-1,-\tfrac12,-\tfrac12,\,-\sqrt{3},\tfrac{\sqrt{3}}{2},-\tfrac{\sqrt{3}}{2}\Bigr)$ \\
$e_{36}$ & $(2\omega^2,\omega^2,\omega)$ &
$\Bigl(-1,-\tfrac12,-\tfrac12,\,-\sqrt{3},-\tfrac{\sqrt{3}}{2},\tfrac{\sqrt{3}}{2}\Bigr)$ \\
$e_{37}$ & $(2\omega,1,1)$ &
$\bigl(-1,1,1,\,\sqrt{3},0,0\bigr)$ \\
$e_{38}$ & $(2\omega,\omega^2,\omega)$ &
$\Bigl(-1,-\tfrac12,-\tfrac12,\,\sqrt{3},-\tfrac{\sqrt{3}}{2},\tfrac{\sqrt{3}}{2}\Bigr)$ \\
$e_{39}$ & $(2\omega,\omega,\omega^2)$ &
$\Bigl(-1,-\tfrac12,-\tfrac12,\,\sqrt{3},\tfrac{\sqrt{3}}{2},-\tfrac{\sqrt{3}}{2}\Bigr)$ \\
$b_{1121}$ & $(2,-1,1)$ &
$\bigl(2,-1,1,\,0,0,0\bigr)$ \\
$b_{1122}$ & $(2,-\omega,\omega^2)$ &
$\Bigl(2,\tfrac12,-\tfrac12,\,0,-\tfrac{\sqrt{3}}{2},-\tfrac{\sqrt{3}}{2}\Bigr)$ \\
$b_{1123}$ & $(2,-\omega^2,\omega)$ &
$\Bigl(2,\tfrac12,-\tfrac12,\,0,\tfrac{\sqrt{3}}{2},\tfrac{\sqrt{3}}{2}\Bigr)$ \\
$b_{1124}$ & $(2,-\omega,\omega)$ &
$\Bigl(2,\tfrac12,-\tfrac12,\,0,-\tfrac{\sqrt{3}}{2},\tfrac{\sqrt{3}}{2}\Bigr)$ \\
$b_{1125}$ & $(2,-\omega^2,1)$ &
$\Bigl(2,\tfrac12,1,\,0,\tfrac{\sqrt{3}}{2},0\Bigr)$ \\
$b_{1126}$ & $(2,-1,\omega^2)$ &
$\Bigl(2,-1,-\tfrac12,\,0,0,-\tfrac{\sqrt{3}}{2}\Bigr)$ \\
$b_{1127}$ & $(2,-\omega^2,\omega^2)$ &
$\Bigl(2,\tfrac12,-\tfrac12,\,0,\tfrac{\sqrt{3}}{2},-\tfrac{\sqrt{3}}{2}\Bigr)$ \\
$b_{1128}$ & $(2,-\omega,1)$ &
$\Bigl(2,\tfrac12,1,\,0,-\tfrac{\sqrt{3}}{2},0\Bigr)$ \\
$b_{1129}$ & $(2,-1,\omega)$ &
$\Bigl(2,-1,-\tfrac12,\,0,0,\tfrac{\sqrt{3}}{2}\Bigr)$ \\
$b_{2121}$ & $(-1,2,1)$ &
$\bigl(-1,2,1,\,0,0,0\bigr)$ \\
$b_{2122}$ & $(-1,2\omega,\omega^2)$ &
$\Bigl(-1,-1,-\tfrac12,\,0,\sqrt{3},-\tfrac{\sqrt{3}}{2}\Bigr)$ \\
$b_{2123}$ & $(-1,2\omega^2,\omega)$ &
$\Bigl(-1,-1,-\tfrac12,\,0,-\sqrt{3},\tfrac{\sqrt{3}}{2}\Bigr)$ \\
$b_{2124}$ & $(-1,2\omega,\omega)$ &
$\Bigl(-1,-1,-\tfrac12,\,0,\sqrt{3},\tfrac{\sqrt{3}}{2}\Bigr)$ \\
$b_{2125}$ & $(-1,2\omega^2,1)$ &
$\bigl(-1,-1,1,\,0,-\sqrt{3},0\bigr)$ \\
$b_{2126}$ & $(-1,2,\omega^2)$ &
$\Bigl(-1,2,-\tfrac12,\,0,0,-\tfrac{\sqrt{3}}{2}\Bigr)$ \\
$b_{2127}$ & $(-1,2\omega^2,\omega^2)$ &
$\Bigl(-1,-1,-\tfrac12,\,0,-\sqrt{3},-\tfrac{\sqrt{3}}{2}\Bigr)$ \\
$b_{2128}$ & $(-1,2\omega,1)$ &
$\bigl(-1,-1,1,\,0,\sqrt{3},0\bigr)$ \\
$b_{2129}$ & $(-1,2,\omega)$ &
$\Bigl(-1,2,-\tfrac12,\,0,0,\tfrac{\sqrt{3}}{2}\Bigr)$ \\
$b_{1131}$ & $(2,1,-1)$ &
$\bigl(2,1,-1,\,0,0,0\bigr)$ \\
$b_{1132}$ & $(2,\omega,-\omega^2)$ &
$\Bigl(2,-\tfrac12,\tfrac12,\,0,\tfrac{\sqrt{3}}{2},\tfrac{\sqrt{3}}{2}\Bigr)$ \\
$b_{1133}$ & $(2,\omega^2,-\omega)$ &
$\Bigl(2,-\tfrac12,\tfrac12,\,0,-\tfrac{\sqrt{3}}{2},-\tfrac{\sqrt{3}}{2}\Bigr)$ \\
$b_{1134}$ & $(2,\omega,-\omega)$ &
$\Bigl(2,-\tfrac12,\tfrac12,\,0,\tfrac{\sqrt{3}}{2},-\tfrac{\sqrt{3}}{2}\Bigr)$ \\
$b_{1135}$ & $(2,\omega^2,-1)$ &
$\Bigl(2,-\tfrac12,-1,\,0,-\tfrac{\sqrt{3}}{2},0\Bigr)$ \\
$b_{1136}$ & $(2,1,-\omega^2)$ &
$\Bigl(2,1,\tfrac12,\,0,0,\tfrac{\sqrt{3}}{2}\Bigr)$ \\
$b_{1137}$ & $(2,\omega^2,-\omega^2)$ &
$\Bigl(2,-\tfrac12,\tfrac12,\,0,-\tfrac{\sqrt{3}}{2},\tfrac{\sqrt{3}}{2}\Bigr)$ \\
$b_{1138}$ & $(2,\omega,-1)$ &
$\Bigl(2,-\tfrac12,-1,\,0,\tfrac{\sqrt{3}}{2},0\Bigr)$ \\
$b_{1139}$ & $(2,1,-\omega)$ &
$\Bigl(2,1,\tfrac12,\,0,0,-\tfrac{\sqrt{3}}{2}\Bigr)$ \\
$b_{3131}$ & $(-1,1,2)$ &
$\bigl(-1,1,2,\,0,0,0\bigr)$ \\
$b_{3132}$ & $(-1,\omega,2\omega^2)$ &
$\Bigl(-1,-\tfrac12,-1,\,0,\tfrac{\sqrt{3}}{2},-\sqrt{3}\Bigr)$ \\
$b_{3133}$ & $(-1,\omega^2,2\omega)$ &
$\Bigl(-1,-\tfrac12,-1,\,0,-\tfrac{\sqrt{3}}{2},\sqrt{3}\Bigr)$ \\
$b_{3134}$ & $(-1,\omega,2\omega)$ &
$\Bigl(-1,-\tfrac12,-1,\,0,\tfrac{\sqrt{3}}{2},\sqrt{3}\Bigr)$ \\
$b_{3135}$ & $(-1,\omega^2,2)$ &
$\Bigl(-1,-\tfrac12,2,\,0,-\tfrac{\sqrt{3}}{2},0\Bigr)$ \\
$b_{3136}$ & $(-1,1,2\omega^2)$ &
$\bigl(-1,1,-1,\,0,0,-\sqrt{3}\bigr)$ \\
$b_{3137}$ & $(-1,\omega^2,2\omega^2)$ &
$\Bigl(-1,-\tfrac12,-1,\,0,-\tfrac{\sqrt{3}}{2},-\sqrt{3}\Bigr)$ \\
$b_{3138}$ & $(-1,\omega,2)$ &
$\Bigl(-1,-\tfrac12,2,\,0,\tfrac{\sqrt{3}}{2},0\Bigr)$ \\
$b_{3139}$ & $(-1,1,2\omega)$ &
$\bigl(-1,1,-1,\,0,0,\sqrt{3}\bigr)$ \\
$b_{2231}$ & $(1,2,-1)$ &
$\bigl(1,2,-1,\,0,0,0\bigr)$ \\
$b_{2232}$ & $(1,2\omega,-\omega^2)$ &
$\Bigl(1,-1,\tfrac12,\,0,\sqrt{3},\tfrac{\sqrt{3}}{2}\Bigr)$ \\
$b_{2233}$ & $(1,2\omega^2,-\omega)$ &
$\Bigl(1,-1,\tfrac12,\,0,-\sqrt{3},-\tfrac{\sqrt{3}}{2}\Bigr)$ \\
$b_{2234}$ & $(1,2\omega,-\omega)$ &
$\Bigl(1,-1,\tfrac12,\,0,\sqrt{3},-\tfrac{\sqrt{3}}{2}\Bigr)$ \\
$b_{2235}$ & $(1,2\omega^2,-1)$ &
$\bigl(1,-1,-1,\,0,-\sqrt{3},0\bigr)$ \\
$b_{2236}$ & $(1,2,-\omega^2)$ &
$\Bigl(1,2,\tfrac12,\,0,0,\tfrac{\sqrt{3}}{2}\Bigr)$ \\
$b_{2237}$ & $(1,2\omega^2,-\omega^2)$ &
$\Bigl(1,-1,\tfrac12,\,0,-\sqrt{3},\tfrac{\sqrt{3}}{2}\Bigr)$ \\
$b_{2238}$ & $(1,2\omega,-1)$ &
$\bigl(1,-1,-1,\,0,\sqrt{3},0\bigr)$ \\
$b_{2239}$ & $(1,2,-\omega)$ &
$\Bigl(1,2,\tfrac12,\,0,0,-\tfrac{\sqrt{3}}{2}\Bigr)$ \\
$b_{3231}$ & $(1,-1,2)$ &
$\bigl(1,-1,2,\,0,0,0\bigr)$ \\
$b_{3232}$ & $(1,-\omega,2\omega^2)$ &
$\Bigl(1,\tfrac12,-1,\,0,-\tfrac{\sqrt{3}}{2},-\sqrt{3}\Bigr)$ \\
$b_{3233}$ & $(1,-\omega^2,2\omega)$ &
$\Bigl(1,\tfrac12,-1,\,0,\tfrac{\sqrt{3}}{2},\sqrt{3}\Bigr)$ \\
$b_{3234}$ & $(1,-\omega,2\omega)$ &
$\Bigl(1,\tfrac12,-1,\,0,-\tfrac{\sqrt{3}}{2},\sqrt{3}\Bigr)$ \\
$b_{3235}$ & $(1,-\omega^2,2)$ &
$\Bigl(1,\tfrac12,2,\,0,\tfrac{\sqrt{3}}{2},0\Bigr)$ \\
$b_{3236}$ & $(1,-1,2\omega^2)$ &
$\bigl(1,-1,-1,\,0,0,-\sqrt{3}\bigr)$ \\
$b_{3237}$ & $(1,-\omega^2,2\omega^2)$ &
$\Bigl(1,\tfrac12,-1,\,0,\tfrac{\sqrt{3}}{2},-\sqrt{3}\Bigr)$ \\
$b_{3238}$ & $(1,-\omega,2)$ &
$\Bigl(1,\tfrac12,2,\,0,-\tfrac{\sqrt{3}}{2},0\Bigr)$ \\
$b_{3239}$ & $(1,-1,2\omega)$ &
$\bigl(1,-1,-1,\,0,0,\sqrt{3}\bigr)$ \\
\end{longtable}
\endgroup

The 165 rays listed in Table~\ref{tab:all-vectors} are defined only up to
overall complex phases.  In order to compare their realifications in
$\mathbb{R}^6$, it is necessary to make a consistent phase choice.  A
direct symbolic computation shows that one can assign to each ray
$v_k\in\mathbb{C}^3$ a phase
\(
  \theta_k = n_k\pi/K
\)
with integers $n_k$ and some fixed (nonunique) integer $K$
(we may take, for instance, $K=1009$), such that the following two conditions
hold simultaneously:
(i) whenever $\langle v_i,v_j\rangle=0$ in $\mathbb{C}^3$, the corresponding
realified vectors $R_i,R_j\in\mathbb{R}^6$ satisfy $R_i\cdot R_j=0$; and
(ii) whenever $\langle v_i,v_j\rangle\neq0$, the realifications satisfy
$R_i\cdot R_j\neq0$.
In particular, this phase choice introduces no spurious orthogonality
relations in $\mathbb{R}^6$ and preserves all of the orthogonality relations
that define the original KS hypergraph in $\mathbb{C}^3$.
We verified the existence of such phases numerically with a simple
backtracking script and confirmed that $K=1009$ suffices.
The existence of such a rational phase assignment demonstrates that the
entire configuration admits an analytic (real-algebraic) realification
without altering its orthogonality structure.

\section{Breakdown of Admissibility Criteria in \texorpdfstring{$\mathbb{R}^6$}{R6}}

Having established the explicit embedding of the 165-ray configuration into $\mathbb{R}^6$, we now turn to its implications for quantum contextuality. This section serves as a concrete illustration of the general fact that faithful real embeddings do not preserve the KS property when contexts are interpreted as maximal in the larger space. We analyse the admissibility criteria for two-valued states on the embedded set and quantify, via numerical optimization, the extent to which the logical contradiction present in $\mathbb{C}^3$ is relaxed in $\mathbb{R}^6$.

The Kochen-Specker contradiction arises from the impossibility of assigning a two-valued state (also called a valuation) $v(P) \in \{0,1\}$ to every projector $P$ in a set $\mathcal{S}$ such that the assignment satisfies two fundamental admissibility criteria:
\begin{enumerate}
    \item \textbf{Exclusivity:} For any pair of mutually orthogonal projectors $P_i, P_j$, the values must satisfy $v(P_i + P_j) = v(P_i) + v(P_j)$. In particular, if $v(P_i)=1$, then $v(P_j)=0$ for all $P_j \perp P_i$.
    \item \textbf{Completeness:} The assignment must respect the identity observable, $v(\mathbb{I}) = 1$.
\end{enumerate}

In the original complex space $\mathbb{C}^3$, any context $\mathcal{C} = \{ P_1, P_2, P_3 \}$ consisting of rank-1 projectors onto an orthonormal basis satisfies $\sum_{i=1}^3 P_i = \mathbb{I}_{\mathbb{C}^3}$. Combining Exclusivity and Completeness yields the strict sum rule:
\begin{equation}
    \sum_{i=1}^3 v(P_i) = v\left( \sum_{i=1}^3 P_i \right) = v(\mathbb{I}_{\mathbb{C}^3}) = 1.
    \label{eq:complex-sum-rule}
\end{equation}
Thus, within every context in $\mathbb{C}^3$, exactly one projector must be assigned the value $1$.

However, under the faithful embedding into $\mathbb{R}^6$, we denote the
projectors mapped from a single complex context by
$\{\Pi_1,\Pi_2,\Pi_3\}$; more generally, we write $\Pi_{i,j}$ for the
projector onto the $i$-th ray in the $j$-th context. The triple
$\{\Pi_1,\Pi_2,\Pi_3\}$ spans only a three-dimensional subspace.
While they remain mutually orthogonal (satisfying Exclusivity), they do not sum to the identity of the embedding space:
\begin{equation}
    \sum_{i=1}^3 \Pi_i = \Pi_{\text{sub}} \neq \mathbb{I}_{\mathbb{R}^6}.
\end{equation}
Completeness applies only to the full space $\mathbb{R}^6$. To invoke it, one must extend the set $\{ \Pi_1, \Pi_2, \Pi_3 \}$ with orthogonal projectors $\{ \Pi_4, \Pi_5, \Pi_6 \}$ completing the basis, such that $\sum_{k=1}^6 \Pi_k = \mathbb{I}_{\mathbb{R}^6}$. The Completeness criterion then imposes:
\begin{equation}
    \sum_{k=1}^6 v(\Pi_k) = 1.
\end{equation}
Restricted to the original embedded context, the strict equality of Eq.~\eqref{eq:complex-sum-rule} relaxes to an inequality:
\begin{equation}
    0 \leq \sum_{i=1}^3 v(\Pi_i) \leq 1.
    \label{eq:relaxed-sum}
\end{equation}
This relaxation permits the assignment $v(\Pi_1)=v(\Pi_2)=v(\Pi_3)=0$ (provided $v(\Pi_k)=1$ for some $k \in \{4,5,6\}$). Consequently, one can assign the value $0$ to all 165 embedded rays in $\mathbb{R}^6$ simultaneously without violating Exclusivity or Completeness, thereby avoiding the KS contradiction.
In $\mathbb{C}^3$ the same ``all-zero'' assignment on the $165$ rays would yield
$\sum_{i=1}^3 v(\Pi_i)=0$ in \emph{every} context, thereby violating the completeness
requirement $v(\mathbb{I}_{\mathbb{C}^3})=1$.

While Eq.~\eqref{eq:relaxed-sum} allows for the local possibility that a specific embedded context sums to 1 (i.e., retains the ``1'' within the embedded subspace), the KS obstruction in $\mathbb{C}^3$ manifests globally in $\mathbb{R}^6$ as a strict bound on the total valuation. Considering the complete set of $|\mathcal{C}| = 130$ contexts in the configuration, as analysed in detail by Navara and Svozil~\cite{svozil-2025-MYOCHS}, if it were possible to assign values such that every embedded context retained the value 1, the total sum would be exactly 130. However, since this is equivalent to a Kochen-Specker coloring in $\mathbb{C}^3$ (which is impossible), the global sum is strictly bounded:
\begin{equation}
     0 \leq \sum_{j=1}^{130} \sum_{i=1}^3 v(\Pi_{i,j}) < 130.
\end{equation}
The lower bound of $0$ corresponds to the ``all-zeros'' assignment mentioned earlier where the value $1$ is placed in the orthogonal complement for every context. The strict upper bound ($<130$) implies that there is a minimum non-zero deficit.
Numerical optimization over all two-valued assignments compatible with Exclusivity and Completeness confirms that at least two contexts must unavoidably surrender the value $1$ to the extra dimensions of $\mathbb{R}^6$ to resolve the logical contradiction. Thus, the final bound is tightened to:
\begin{equation}
    0 \leq \sum_{j=1}^{130} \sum_{i=1}^3 v(\Pi_{i,j}) \leq 128.
\end{equation}

\section{Discussion}

We have described a simple phase-adjusted realification procedure that embeds any finite set of rays in $\mathbb{C}^3$ into $\mathbb{R}^6$ in a way that preserves and reflects complex orthogonality. The construction is purely algebraic and uses only the phase freedom inherent in projective rays. Applied to the 165-ray configuration built from MUBs, it yields a faithful orthogonal representation in $\mathbb{R}^6$ (after suitable phase choices), and, by trivial extension, in $\mathbb{R}^7$ and above.

Table~\ref{tab:all-vectors} lists all 165 rays explicitly in $\mathbb{C}^3$ together with their canonical realifications in $\mathbb{R}^6$. For applications requiring strict faithfulness (no extra orthogonality among non-orthogonal rays), one additionally picks phases $\theta_k$ as in Sec.~II before applying $\Phi_0$, which simply rotates each pair $(\Re z_j,\Im z_j)$ in its two-dimensional real subspace.

Whereas, in its original complex realisation in $\mathbb{C}^3$, the 165-ray,
130-context configuration is KS uncolourable (there is no
two-valued state assigning $0$ and $1$ such that in each 3-element context
exactly one projector has value $1$~\cite{Cabello2025}), the same set of rays, when embedded
in $\mathbb{R}^6$, does admit classical two-valued states if we keep only
the embedded 165 rays and the original 3-element contexts.

The reason is that, in $\mathbb{R}^6$, each such triple now spans only a
3-dimensional subspace and is no longer a maximal context: it can always be
completed to a full orthonormal basis of $\mathbb{R}^6$ by adding three extra
vectors not belonging to our configuration. A valuation that assigns the
value $0$ to all 165 given rays, and places the value $1$ on one of these
additional basis vectors for each completed context, then satisfies all
Kochen-Specker constraints on the restricted set. Because there is a
continuum of choices for the orthonormal complements (each 3-dimensional
subspace has infinitely many orthonormal bases of its orthogonal complement),
one can arrange the completions so that no consistency problems arise across
different contexts.

This contrast highlights that a faithful orthogonal representation of a KS
hypergraph in a higher-dimensional real space does not automatically preserve
KS uncolourability: classical two-valued interpretations exist for the
embedded configuration in $\mathbb{R}^6$, whereas they are forbidden for its
realisation as maximal contexts in $\mathbb{C}^3$.

There are two conceptual messages that emerge from this analysis.
First, our construction recovers and reformulates, in explicitly MUB-based
terms, the central conclusion of Harding and Salinas Schmeis
\cite{harding2025remarksIJTP}: in three dimensions there exist finite
orthogonality configurations of rank-1 projectors that are representable in
$\mathbb{C}^3$ but admit no faithful orthogonal representation~\cite{lovasz-79} in
$\mathbb{R}^3$. In particular, the triple Yu--Oh construction
\cite{Yu-2012,Cabello2025}, which can be implemented experimentally for a
single qutrit (for instance via generalized multiport interferometers
\cite{rzbb}), realises such a configuration: its hypergraph of orthogonality
relations can be realised by unit vectors in $\mathbb{C}^3$ but not by unit
vectors in $\mathbb{R}^3$.

Second, the same triple Yu--Oh configuration, when completed to a
Kochen-Specker set in $\mathbb{C}^3$, is KS-uncolourable: there is no
two-valued state (global $0$--$1$ assignment) compatible with all its
3-element contexts. Our phase-adjusted embedding shows that this
configuration does admit a faithful orthogonal representation in
$\mathbb{R}^6$ (and trivially in $\mathbb{R}^D$ for all $D\ge 6$). However,
in $\mathbb{R}^6$ each 3-element context spans only a 3-dimensional subspace
and is no longer maximal; it can always be extended to a full orthonormal
basis of $\mathbb{R}^6$ by adding three extra vectors outside the given set.
One can then construct classical two-valued states by assigning the value
$0$ to all 165 embedded rays and the value $1$ to one of the additional
basis vectors in each completed context. Because each 3-dimensional subspace
has a continuum of orthogonal complements, the completions can be chosen
consistently across contexts. Thus, the complex realisation in $\mathbb{C}^3$
is inherently KS-nonclassical, whereas the real realisation in
$\mathbb{R}^6$ admits classical hidden-variable (two-valued state) models once the
contexts are interpreted as maximal in the larger space.

This sharpens and complements earlier tests that distinguish real and
complex quantum theory at the level of measurement statistics
\cite{mckague-2009,renou-2021,wu-2022}: here the separation appears already
at the level of the logical structure, in the existence or absence of
two-valued states on the same finite orthogonality hypergraph when realised
in $\mathbb{C}^3$ versus in $\mathbb{R}^6$.

\begin{acknowledgments}
The authors thank Felix Effenberger for communications related to the subject.

This text was partially created and revised with assistance from one or more of the following large language models:  Gemini (2.5 Pro as well as 3.0 Pro Preview)
as well as gpt-5.1-high. All content, ideas, and prompts were provided by the authors.

AK was supported by JST, CREST Grant Number JPMJCR23P4, Japan.
KS was funded by the Austrian Science Fund (FWF) Grant \href{https://doi.org/10.55776/PIN5424624}{10.55776/PIN5424624}.
The authors acknowledge TU Wien Bibliothek for financial support through its Open Access Funding Programme.

\end{acknowledgments}

\bibliography{svozil}

\end{document}